\documentclass[pra,aps,longbibliography,twocolumn,superscriptaddress,floatfix,footnoteinbib]{revtex4-1}
\usepackage{amsmath,amssymb,graphicx,units}
\usepackage[usenames,dvipsnames]{color}
\usepackage{verbatim}
\usepackage{lipsum}
\usepackage{setspace}
\usepackage{dcolumn}
\usepackage{bm}
\usepackage[bookmarks=false,linkcolor=blue,urlcolor=blue,colorlinks,citecolor=blue]{hyperref}
\renewcommand{\vec}[1]{\boldsymbol{#1}}
\usepackage{mathrsfs}
\usepackage[T1]{fontenc}

\begin{document}
	
\title{Optimal Dynamical Gauge in the Quantum Rabi Model}

\author{Yuqi Qing}
\affiliation{State Key Laboratory of Information Photonics and Optical Communications, Beijing University of Posts and Telecommunications, Beijing 100876, China}

\author{Wen-Long You}
\email{wlyou@nuaa.edu.cn}
\affiliation{College of Science, Nanjing University of Aeronautics and Astronautics, Nanjing, 211106, China}

\author{Yueheng Lan}
\email{lanyh@bupt.edu.cn}
\affiliation{State Key Laboratory of Information Photonics and Optical Communications, Beijing University of Posts and Telecommunications, Beijing 100876, China}

\author{Maoxin Liu}
\email{mxliu@bnu.edu.cn}
\affiliation{School of Systems Science/Institute of Nonequilibrium Systems, Beijing Normal University, Beijing 100875, China}

\begin{abstract}
In this paper, we investigate the gauge dependence of various physical observables in the quantum Rabi model (QRM) under different potential fields, arising from the Hilbert-space truncation of the atomic degree of freedom. We discover that in both the square-well potential and oscillator potential, the optimal gauges for the ground-state energy of the QRM vary with respect to the cavity frequency, with the dipole gauge being optimal in the low-frequency limit and the Coulomb gauge in the high-frequency limit of the cavity frequency. Additionally, for higher energy levels, the optimal gauge asymptotically approaches the dipole gauge. However, for the dynamical quantity out-time-order correlator (OTOC), we find the necessity to introduce an optimal dynamical gauge. We determine the optimal dynamical gauge by minimizing the mean error between the two-level OTOC and the full Hamiltonian one. We expect that this study will contribute to a more profound understanding of the subtle relation between gauge choice and the dynamics of QED systems.
\end{abstract}

\maketitle


\section{Introduction\label{sec1}}

The concept of gauge is the cornerstone of modern physics. In classical electromagnetism, the observables, such as the electric field strength $\vec{E}$ and the magnetic induction strength $\vec{B}$, remain invariant under a gauge transformation, known as gauge invariance. Gauge invariance holds even more significance in gauge field theory, which forms the basis of the Standard Model \cite{cabibbo2017introduction}. In non-relativistic quantum electrodynamics (QED) theory, the gauge invariance allows for the derivation of an equivalent set of interaction forms \cite{cohen1997photons}. A prominent example is the equivalence between the $\vec{p} \cdot \vec{A}$ interaction in the Coulomb gauge and the $\vec{r} \cdot \vec{E}$ interaction in the dipole gauge, which can be demonstrated through a unitary transformation dubbed the Goppert-Mayer (GM) transformation \cite{goppert1931elementarakte}. However, there are instances that different gauges can result in diverse physical results, and certain gauges may lead to non-physical predictions. This issue is commonly referred to as gauge dependence. For instance, Lamb, in his seminal work on the fine structure of hydrogen \cite{lamb1952fine}, pointed out that while different representations of interaction in the electric field-dipole system yielded identical results, the probability amplitudes can only be correctly interpreted in the dipole gauge rather than the Coulomb gauge. Furthermore, Power and Zienau proposed in their work \cite{power1959coulomb} that in the Coulomb gauge the presence of non-physical precursors is associated with singularities at the origin in certain momentum space representations of the propagators. The singularities can be eliminated by performing the Power-Zienau-Woolley transformation from the Coulomb gauge to the multipolar gauge \cite{woolley1971molecular}. Furthermore, a study on two-photon transitions \cite{bassani1977choice} demonstrated the superiority of the dipole gauge in calculations, as it required  fewer intermediate states for computation. It was also revealed that the gauge dependence is a recurring phenomenon and has undergone frequent examination in various studies \cite{forney1977choice, kobe1978gauge, kobe1978question, yang1982gauge, lamb1987matter, woolley2000gauge, rzkazewski2004equivalence}.

In the context of light-matter interaction, the quantum Rabi model (QRM) has long served as a paradigmatic model for describing the interaction between a two-level atom and an optical cavity field \cite{rabi1936process, rabi1937space}. Recently, the so-called ultra-strong coupling (USC) and deep strong coupling (DSC) have been achieved in solid state quantum systems \cite{casanova2010deep, yoshihara2017superconducting, yoshihara2018inversion, mueller2020deep}, triggering renewed experimental and theoretical research interests, which encompass investigations into coupling induced effects \cite{de2014light, le2016fate}, nonlinear optics with higher-order processes \cite{stassi2017quantum}, and the ability to regulate chemical properties \cite{galego2019cavity}. The QRM is considered to effectively describe the behavior of a flux qubit coupled to an LC resonator, which is the basic unit of superconducting circuits \cite{romero2012ultrafast, stassi2018long, macri2018simple, forn2019ultrastrong, stassi2020scalable}. The QRM has been used to construct artificial gauge field systems, providing an ideal platform for manipulating two-level systems and engineering topological states of matter \cite{forn2016broken, gely2017convergence, manucharyan2017resilience, galitski2019artificial, zhang2021quantum, padilla2022understanding}. Due to the subtle nature of quantum few-body interactions, a superradiant phase transition in the QRM has been predicted \cite{hepp1973superradiant, rzazewski1975phase, hwang2015quantum, hwang2016quantum, liu2017universal} and subsequently observed in experiments \cite{guerci2020superradiant, chen2021experimental, zheng2023observation}. Lately, Braak made a remarkable breakthrough by obtaining the eigenenergy spectrum of the QRM \cite{braak2011integrability}, which sheds light on subsequent studies on the integrability of quantum systems \cite{larson2013integrability, he2015exact, grimaudo2023quantum}.

Recent researches have pointed out that the QRM, due to the two-level truncation applied to the atomic Hilbert space in the derivation of the Hamiltonian, is actually gauge dependent \cite{2018Breakdown,stokes2019gauge}. It is noted that this gauge dependence becomes more pronounced in the USC regime and the DSC regime \cite{2018Breakdown}. When specific potentials are chosen, it has been demonstrated that, while the two-level approximation performed in the dipole gauge can be valid, the Coulomb gauge manifestly fails in the USC regime. In Ref. \cite{stokes2019gauge}, it was suggested that a Jaynes-Cummings model (JCM) could be derived from a full Hamiltonian directly by choosing a specific gauge, namely the Jaynes-Cummings (JC) gauge, without needing a rotating wave approximation. Moreover, the JC gauge yields more accurate energy spectra than those of the QRM, thereby dispelling the preconception that the JCM is less fundamental than the QRM in the USC regime. Naturally, instead of pursuing the optimal gauge  on a case-by-case basis, it is more advanced to develop a gauge-independent truncation scheme in the light-matter interaction Hamiltonian. A recent work highlights that the gauge dependence arises from non-localities introduced during the construction of the effective Hamiltonian in the truncated Hilbert space \cite{di2019resolution}. Thereby, one approach to compensate for non-localities in the atomic potential involves implementing a minimal-coupling replacement in the non-local potential and truncating both the atomic Hamiltonian and the gauge transformation unitary operator \cite{di2019resolution, savasta2021gauge}. An alternative approach emphasizes the substitution of truncated operators in each term of the full Hamiltonian, resulting in a consistent definition of these operators within the truncated Hilbert space \cite{taylor2020resolution, taylor2022resolving, ryu2023matrix}.In a recent review \cite{stokes2022implications}, the authors propose an interesting perspective suggesting that the definition of QED subsystems is linked to the choice of gauge. Nevertheless, the investigation on gauge dependence has enlighten and broadened studies on gauge-independent interaction theory, including operational measurements \cite{settineri2021gauge}, emission spectra \cite{mercurio2022regimes}, open systems \cite{salmon2022gauge}, arbitrary media \cite{gustin2023gauge}, Dicke model \cite{garziano2020gauge,akbari2023generalized} and Hopfield model \cite{garziano2020gauge}.

Recently, the dependence of the optimal gauge choice on multi-cavity modes has been studied \cite{roth2019optimal,arwas2023properties}. It was found that the optimal gauge generally depends on the modes and observables. These findings inspire the recognition that, before fully addressing the issue of gauge dependence, it is crucial to undertake in-depth and specialized research on the gauge ambiguity. In this work, we examine various forms of the external potentials acting on the atom, aiming to quantitatively illustrate the gauge dependence by investigating observables under the consideration, such as the ground-state energy, energy spectra, dynamics, etc.

The structure of this paper is organized as follows. In Sec. \ref{sec2}, we revisit the derivation of the QRM in an arbitrary gauge formalism with a finite-level truncation. We lay out the theoretical framework and necessary background for the coming numerical calculations. In Sec.\ref{sec3}, we perform numerical calculations of various quantities of interest in the QRM with different cavity frequencies and potentials. By comparing these results with the accurate results obtained from the full Hamiltonian, we identify both the optimal static gauges and optimal dynamical gauge and consider their dependence on the potentials and cavity frequencies. This analysis will shed light on the gauge dependence and provide insights into the optimal gauge choices in different scenarios. Finally, in Sec.\ref{sec4}, we present a concise summary of our findings and discuss the implications of the optimal gauge. We highlight the significance of understanding the gauge dependence in the QRM and its potential applications in resolving gauge ambiguity and developing improved approximate methods in practical calculations.

\section{Quantum Rabi model in an arbitrary-gauge formalism with a finite-level truncation\label{sec2}}

To investigate the role of gauge, we should embody the gauge freedom in the Hamiltonian explicitly. In this section, we propose a general framework for deriving the atom-field interaction Hamiltonian in an arbitrary-gauge formalism with a finite-level truncation. Based on this framework, we can conveniently study the influence of gauge on different physical observables in the truncated Hilbert space.

\subsection{The minimal-coupling Lagrangian and Hamiltonian in the Coulomb gauge}

We start the construction of the QRM Hamiltonian in an arbitrary-gauge formalism from a minimal-coupling Lagrangian form. An atom in the electromagnetic field and an external potential can be formulated with this minimal-coupling approach, can be formulated as a summation of the atomic part, the field part and their interaction
\begin{align}
	L = L_{a}+L_{f}+L_{i} \label{L_minimal_coupling},
\end{align}
where the atomic, the field
and atom-field interaction Lagrangians read 
\begin{align}
	L _{a}&= \frac{1}{2} m \dot{\vec{r}}^2-V_{\text{ext}}(\vec{r}) \label{L_a}, \\
	L _{f}&=  \frac{\varepsilon_0}{2} \int d^3x \left[(-\dot{\vec{A}}(\vec{x}) - \nabla \phi(\vec{x})) ^2-  (\nabla \times \vec{A}(\vec{x}))^2  \right] \label{L_f}, \\
	L_{i}&= - q\phi(\vec{r}) + q\dot{\vec{r}} \cdot \vec{A}(\vec{r}) \label{L_i}.
\end{align}
Here $m$, $q$, $\vec{r}$ are the reduced mass, charge and the position of the atom. $V_{\rm ext}(\vec{r})$ is the external potential that confines the atom. For convenience, we set the reduced Planck constant and the speed of light to unity, i.e., $\hbar=c=1$. In order to distinguish from atom position $\vec{r}$, we use $\vec{x}$ denoting the coordinate of the field. $\varepsilon_0$ is the permittivity of free space. $\phi$ and $\vec{A}$ are the electromagnetic scalar potential and the vector potential, satisfying 
\begin{align}
	\vec{E}(\vec{x},t) &= -\dot{\vec{A}}(\vec{x},t) - \nabla \phi(\vec{x},t) \label{E}, \\
	\vec{B}(\vec{x},t) &= \nabla \times \vec{A}(\vec{x},t) \label{B},
\end{align}
where $\vec{E}$ and $\vec{B}$ are the electric field strength and magnetic induction, respectively. Throughout this paper, in most cases, the explicit dependence on time $t$ is ignored for the sake of simplicity in writing. The interaction part of the Lagrangian readily leads to a Lorentz invariant action
$S_{\rm int}= -\int qA^{\mu}dx_{\mu}$, where $A^{\mu} = (\phi, \vec{A})$ and $x_{\mu}=(t,-\vec{x})$ are four-vectors.

Although the minimal-coupling Lagrangian $L$ in Eq. \eqref{L_minimal_coupling} is already formulated, the gauge of the system has not yet been fixed, resulting in the arbitrariness in the form of vector potential $\vec{A}$. The arbitrariness can be intuitively understood by the so-called Helmholtz decomposition, such that
\begin{align}
	\vec{A}=\vec{A}_\parallel+\vec{A}_\perp \label{Helmholtz_decomposition},
\end{align}
where $\vec{A}_\parallel$ and $\vec{A}_\perp$ are the  the longitudinal  and transverse components of $\vec{A}$ with $\nabla \times \vec{A}_\parallel = 0$ and  $\nabla \cdot \vec{A}_\perp = 0$. The transverse field $\vec{A}_\perp$ is uniquely determined by the dynamic variables of the electromagnetic fields, i.e., $\vec{E}$ and $\vec{B}$, while the longitudinal field $\vec{A}_\parallel$ is a non-physical variable, which can be arbitrarily identified.

We first present the Coulomb gauge, namely, $\vec{A}_\parallel=0$. Note that other gauge forms discussed in this paper can be derived from the Coulomb gauge via a gauge transformation. Within the Coulomb gauge, the scalar potential $\phi$ is actually the static Coulomb potential $V_{\rm coul}$. The vector potential $\vec{A}$ contains only the transverse component $\vec{A}_\perp$, denoted by $\vec{A}^C$. The Lagrangian in the Coulomb gauge can be obtained (see the appendix \ref{appendixA} for a detailed derivation)
\begin{align}
	L^C &= \frac{1}{2} n \dot{\vec{r}}^2 - V(\vec{r}) + q \dot{\vec{r}} \cdot \vec{A}_\perp(\vec{r}) \notag \\
	&+ \frac{\varepsilon_0}{2} \int [\dot{\vec{A}}_\perp^2(\vec{x}) -  (\nabla \times \vec{A}_\perp(\vec{x}))^2] d^3x \label{L_C}.
\end{align}
Here $V(\vec{r})=V_{\rm coul}+V_{\rm ext}(\vec{r})$ is the sum of the Coulomb potential $V_{\rm coul}$ and the external potential $V_{\rm ext}(\vec{r})$. In principle, we can let $V(\vec{r})$ can be an arbitrary potential by adjusting $V_{\rm ext}(\vec{r})$. Taking the position of the atomic position $\vec{r}$ and the transverse field $\vec{A}_\perp$ as canonical coordinates of the atom and the field, corresponding canonical momenta can be defined as
\begin{align}
	& \vec{p}^C = \frac{\partial L^C}{\partial \dot{\vec{r}}} = n\dot{\vec{r}} + q\vec{A}_\perp(\vec{r}) \label{p_C}, \\
	& \vec{\Pi}^C(\vec{x}) = \frac{\delta \mathcal{L}^C}{\delta \dot{\vec{A}}_\perp} = \varepsilon_0 \dot{\vec{A}}_\perp(\vec{x}) \label{pi_C},
\end{align}
where the Lagrangian density is denoted $\mathcal{L}^C$. Implementing a Legendre transformation, we obtain the Hamiltonian in the Coulomb gauge
\begin{align}
	H^C &= \dot{\vec{r}} \cdot \vec{p}^C + \int (\dot{\vec{A}}_\perp(\vec{x}) \cdot \vec{\Pi}^C(\vec{x})) d^3 x - L^C \notag \\
	&= \frac{1}{2m}(\vec{p}^C - q\vec{A}_\perp(\vec{r}))^2 + V(\vec{r}) \notag \\
	&+\frac{\varepsilon_0}{2} \int [(\frac{\vec{\Pi}^C(\vec{x})}{\varepsilon_0})^2 +  (\nabla \times \vec{A}_\perp(\vec{x}))^2 ]d^3 x \label{H_C}.
\end{align}
With all of these equations in hand, one can readily establish a quantum theory by introducing the canonical commutation relation
\begin{align}
	& [\hat{r}_i, \hat{p}_j^C] = i \delta_{i,j} \label{canonical_commutation_relation_rp}, \\
	& [\hat{A}_{\perp,i}(\vec{x}), \hat{\Pi}_{\perp,j}^C(\vec{x}')] = i \delta_{i,j}^{\perp}(\vec{x}-\vec{x}') \label{canonical_commutation_relation_Api},
\end{align}
where subscripts $i$ and $j$ denote three-dimensional components of vectors. 

\subsection{Equivalent theoretical description in an arbitrary-gauge formalism}
The gauge transformation associated with an arbitrary scalar field $\chi(\vec{x},t)$ is given by
\begin{align}
	\phi^{\chi}(\vec{x},t) = \phi^C(\vec{x},t) - \frac{\partial}{\partial t} \chi(\vec{x},t) \label{gauge_transformation_ahi}, \\
	\vec{A}^{\chi}(\vec{x},t) = \vec{A}^C(\vec{x},t) + \nabla \chi(\vec{x},t) \label{gauge_transformation_A}.
\end{align}
The corresponding interaction Lagrangian under this gauge transformation becomes
\begin{align}
	L_{i}^\chi = -q\phi^\chi(\vec{r},t) + q\dot{\vec{r}} \cdot \vec{A}^\chi(\vec{r},t) = L_{i}^C + \frac{d}{dt}[q\chi(\vec{r},t)] \label{L_chi} .
\end{align}
The gauge transformation results in an additional total time derivative term in Eq.\eqref{L_chi}, which makes no contribution to the equations of motion. Thus, the gauge transformation leads to an equivalent transformed Lagrangian
\begin{align}
	L^{\chi}(\phi^C,\vec{A}^C)=L^C(\phi^{\chi},\vec{A}^{\chi})= L^C(\phi,\vec{A})+\frac{d}{dt} [q\chi(\vec{r},t)] \label{L_chi_L_C} .
\end{align}
The canonical momentum of the atom can be then calculated as
\begin{align}
	\vec{p}^{\chi} = \frac{\partial L_{\chi}}{\partial \dot{\vec{r}}} = \vec{p}^C + q\nabla_{\vec{r}} \chi(\vec{r},t) \label{p_chi} .
\end{align}
The canonical momentum of the field also changes like 
\begin{align}
	\vec{\Pi}^{\chi}(\vec{x}) = \frac{\delta \mathcal{L}^{\chi}}{\delta \dot{\vec{A}}_\perp} = \vec{\Pi}^C + q\nabla_{\vec{A}_\perp} \chi(\vec{A}_\perp,t) \label{pi_chi}.
\end{align}
If $\chi$ is explicitly dependent on $\vec{A}_\perp$. Note that the divergence operator $\nabla_{\vec{A}_\perp}$ denotes the derivative of ${\vec{A}_\perp}$. Eq.\eqref{p_chi} can also be achieved via a unitary transformation define by $U^\chi=\exp(iq\chi(\vec{r},t))$. It is the same to define a unitary transformation $U^\chi = \exp(iq\chi(\vec{A}_\perp,t))$.

Following Ref.\cite{stokes2019gauge}, we consider the arbitrary gauge formalism, in which $U^\alpha=\exp(-i\alpha \vec{r} \cdot \vec{A}_\perp)$ is introduced to act on the Coulomb gauge, where $\alpha$ is a real number. Here $\alpha=0$ remains the Coulomb gauge while $\alpha=1$ corresponds to the dipole gauge. The unitary transformation denoted by $U^1=\exp(-i\vec{r}\cdot\vec{A}_\perp)$ is actually the GM transformation \cite{goppert1931elementarakte}. The GM transformation yields dipole interaction form, which only contains gauge-independent quantities $\vec{r}$ and $\vec{E}$ rather than the gauge-dependent interaction term in Eq.\eqref{H_C}. Consequently, in the context of atomic, molecular and optical physics, the dipole interaction form is usually applied. Under the arbitrary-gauge unitary transformation, the canonical momenta transform as
\begin{align}
	\vec{p}^\alpha &= U^{\alpha \dagger} \vec{p}^C U^\alpha = \vec{p}^C - \alpha q\vec{A}_\perp(0) \label{p_alpha_after_U_alpha_transformation}, \\
	\vec{\Pi}^\alpha (\vec{x}) &= U^{\alpha \dagger} \vec{\Pi}^C(\vec{x}) U^\alpha = \vec{\Pi}^C(\vec{x}) - \alpha q\vec{r} \label{pi_alpha_after_U_alpha_transformation} .
\end{align}
It is clear that, if $\alpha = 0$, the canonical momenta restores the ones in the Coulomb gauge. Expressed by the canonical momenta \eqref{p_alpha_after_U_alpha_transformation} and \eqref{pi_alpha_after_U_alpha_transformation}, the Hamiltonian can be written in the arbitrary-gauge form
\begin{align}
	H &= \frac{1}{2m} \left[ \vec{p}^\alpha - (1-\alpha) q\vec{A}_\perp(\vec{r})\right] ^2 + V(\vec{r}) \notag \\
	&+ \frac{\varepsilon_0}{2} \int \left\{ \left[\frac{\vec{\Pi}_{\alpha}(\vec{x})+\alpha q\vec{r}}{\varepsilon_0}\right]^2 + \left[\nabla \times \vec{A}_\perp(\vec{x})\right]^2 \right\} d^3 x \label{full_H_a}.
 \end{align}
For clarity, we partition the Hamiltonian into three parts: the atomic Hamiltonian, the field Hamiltonian, and the interaction Hamiltonian
\begin{align}
	H &= H^{\alpha}_a + H^{\alpha}_f + H^{\alpha}_i, \label{H_by_part}
\end{align}
where
\begin{align}
    H^{\alpha}_a &= \frac{{\vec{p}^\alpha}^2}{2m}  + V(\vec{r}) \label{H_a}, \\
    H^{\alpha}_f &= \frac{\varepsilon_0}{2} \int \left\{\left[\frac{\vec{\Pi}^\alpha(\vec{x})}{\varepsilon_0}\right]^2 + \left[\nabla \times \vec{A}_\perp(\vec{x})\right]^2\right\}d^3x \label{H_f}, \\
    H^{\alpha}_i &= -\frac{(1-\alpha)}{n}q\vec{p}^{\alpha}\vec{A}_\perp(\vec{r}) + \frac{(1-\alpha)^2}{2m}q^2\vec{A}_\perp^2(\vec{r}) \notag \\
    &+ \int [ \frac{\alpha}{\varepsilon_0}q\vec{r} \cdot \vec{\Pi}^\alpha(\vec{x}) + \frac{\alpha^2}{2\varepsilon_0} q^2 \vec{r}^2]d^3x \label{H_i}.
\end{align}

\subsection{Arbitrary-gauge formalism Hamiltonian in a truncated Hilbert space}
In this subsection, we express the Hamiltonian \eqref{full_H_a} in the Fock state representation of the atomic excitation and photon. The desired QRM in the arbitrary-gauge formalism with a finite-level truncation can be easily obtained.

To truncate the Hilbert subspace of the atom, we define an $d$-level projection operator $\mathscr{P}_d^\alpha = \sum_{i=0}^{d-1} |\epsilon_i^\alpha\rangle \langle \epsilon_i^\alpha|$, where $|\epsilon_i^\alpha\rangle$ is the eigenstate of $H_a^\alpha$, satisfying $H_a^\alpha |\epsilon_i^\alpha \rangle = \epsilon_i |\epsilon_i^\alpha \rangle$. In the $d$-level representation of $\{ |\varepsilon_0^\alpha \rangle,|\epsilon_1^\alpha \rangle,|\epsilon_2^\alpha \rangle \cdots |\epsilon_{d-1}^\alpha \rangle \}$, the position, momentum and Hamiltonian operators of atom in Eqs.\eqref{H_a},\eqref{H_f} and \eqref{H_i} can be truncated via
\begin{align}
	\mathscr{P}_d^\alpha \vec{r} \mathscr{P}_d^\alpha & = \sum_{i,j=0}^{d-1} r_{i,j} |\epsilon_i^\alpha \rangle \langle \epsilon_j^\alpha| \label{PrP}, \\
	\mathscr{P}_d^\alpha \vec{p}^\alpha \mathscr{P}_d^\alpha & = \sum_{i,j=0}^{d-1} p^\alpha_{i,j} |\epsilon_i^\alpha \rangle \langle \epsilon_j^\alpha| \label{PpP}, \\
	\mathscr{P}_d^\alpha \vec{H}_a^\alpha \mathscr{P}_d^\alpha & = \sum_{i=0}^{d-1} \epsilon_i |\epsilon_i^\alpha \rangle \langle \epsilon_i^\alpha|, \label{PHpP}
\end{align}
where $r_{i,j}=\langle \epsilon_i^\alpha | \vec{r} | \epsilon_j^\alpha \rangle$, $p_{i,j}^\alpha=\langle \epsilon_i^\alpha | \vec{p}^\alpha | \epsilon_j^\alpha \rangle$. However, the truncation of the term including $\vec{r}^2$ is subtle. We actually take $(\mathscr{P}_d^\alpha \vec{r} \mathscr{P}_d^\alpha)^2$ rather than $\mathscr{P}_d^\alpha \vec{r}^2 \mathscr{P}_d^\alpha$, and this method is widely used by many researches \cite{di2019resolution, stokes2019gauge, taylor2020resolution}.

As for the field Hamiltonian, we first define an operator $a_{\alpha,\lambda}^\dagger(\vec{k}) = \sqrt{\frac{1}{2\omega }}(\omega\vec{\tilde{A}}_{\perp,\lambda}(\vec{k}) + i\tilde{\Pi}_{\perp,\lambda}^\alpha(\vec{k}))$, where $\vec{\tilde{f}}(\vec{k})$ is the Fourier transformation of $\vec{f}(\vec{x})$. $\omega=|\vec{k}|$ is the mode frequency and $\lambda$ denotes the polarization direction. It can be verified that $[a^\dagger_{\alpha, \lambda}(\vec{k}), a_{\alpha, \lambda'}(\vec{k}')] = \delta_{\lambda,\lambda'}\delta(\vec{k}-\vec{k}')$, thus implying that $a_{\alpha,\lambda}(\vec{k}), a_{\alpha,\lambda}^\dagger(\vec{k})$ are actually the creation and annihilation operators of photon mode $(\lambda, \vec{k})$. The field $\vec{A}_\perp(\vec{x})$ and $\vec{\Pi}_\perp^\alpha(\vec{x})$ can be expanded by different modes via a Fourier transformation
\begin{align}
	\vec{A}_\perp(\vec{x}) &= \int d^3 k \sum_\lambda g\vec{e}_\lambda(\vec{k}) a^\dagger_{\alpha, \lambda}(\vec{k}) e^{-i\vec{k}\cdot \vec{x}} + \rm{H.c.} \label{A_expand}, \\
	\vec{\Pi}_{\alpha}(\vec{x}) &= i\int d^3 k \sum_\lambda \omega g\vec{e}_\lambda(\vec{k}) a^\dagger_{\alpha, \lambda}(\vec{k}) e^{-i\vec{k}\cdot \vec{x}} - \rm{H.c}. \label{pi_expand},
\end{align}
where $g=1/\sqrt{2\omega (2\pi)^3}$ is the normalization coefficient used to ensure that the commutation relation \eqref{canonical_commutation_relation_Api} between $\vec{A}_\perp(\vec{x})$ and $\vec{\Pi}^\alpha(\vec{x})$  is satisfied, $\vec{e}_\lambda$ denotes the unit vector in the polarisation direction denoted by $\lambda$. $\rm{H.c.}$ denotes the Hermitian conjugate term. To make a single-mode approximation that preserves the existing gauge invariance, Eqs.\eqref{A_expand} and \eqref{pi_expand} can be simplified to
\begin{align}
	\vec{A}_\perp(\vec{x}) &= g\vec{e}_\lambda(a^\dagger_\alpha + a_\alpha) \label{A_single_mode}, \\
	\vec{\Pi}^\alpha(\vec{x}) &= i\omega g\vec{e}_\lambda(a^\dagger_\alpha - a_\alpha) \label{pi_single_mode},
\end{align}
where $g = 1/\sqrt{2\omega v}$ and $v$ represents the volume of the cavity, which  comes from a finite volume assumption for the mode discretization in the single mode approximation .This choice of $g$ ensures that the commutation relation
\begin{align}
    [A_{\perp,i}(\vec{x}), \Pi_j^\alpha(\vec{x})] = \frac{ie_i e_j}{v}
\end{align}
is satisfied. Here, $e_i$ and $e_j$ are the components of the electric field vector, and $\omega$ corresponds to the frequency. $e_i$ and $e_j$ are Cartesian components of $\vec{e}_\lambda$. The field Hamiltonian after the single mode approximation can be written as $H_f^\alpha= \omega (a^\dagger_\alpha a_\alpha + \frac{1}{2})$. We group the $\vec{A}_\perp^2(\vec{r})$ term into $H_f^\alpha$, and define a pair of bosonic operators through the Bogoliubov transformation
\begin{align}
	H_{f}^{\alpha \prime} = H_{f}^\alpha + \frac{(1-\alpha)^2}{2m}q^2\vec{A}_\perp^2(\vec{r}) =  \omega_\alpha (b_\alpha^\dagger b_\alpha + \frac{1}{2}) \label{Hnf_alpha_A2},
\end{align}
where $\omega_\alpha = \sqrt{\omega^2+(1-\alpha)^2q^2/mv\varepsilon_0}$ is the renormalized frequency. In terms of the bosonic operators, we have
\begin{align}
	\vec{A}_\perp(\vec{x}) &= g_\alpha \vec{e}_\lambda(b^\dagger_\alpha + b_\alpha) \label{A_expressed_by_b}, \\
	\vec{\Pi}^\alpha(\vec{x}) &= i\omega_\alpha g_\alpha\vec{e}_\lambda(b^\dagger_\alpha + b_\alpha) \label{Pi_expressed_by_b}.
\end{align}
Here $g$ is replaced with $g_\alpha = \sqrt{1/(2\omega_\alpha v)}$ to preserve the commutation relation \eqref{new_commutation_relation} between \eqref{A_expressed_by_b} and \eqref{Pi_expressed_by_b}.

Finally, the QRM in the arbitrary-gauge formalism with the $d$-level truncation reads
\begin{align}
	H^{\alpha}_{d} &= \mathscr{P}_d^\alpha H \mathscr{P}_d^\alpha \notag \\
	&= \sum_{i=0}^{d-1} \epsilon_i |\epsilon_i^\alpha \rangle \langle \epsilon_i^\alpha| +  \omega_\alpha (b_\alpha^\dagger b_\alpha + \frac{1}{2}) + \frac{\alpha^2 q^2}{2v\varepsilon_0}\sum_{i,j,k=0}^{d-1} r_{i,k}r_{k,j} \notag \\
	&- \frac{(1-\alpha)q}{n} \sqrt{\frac{1}{2\omega_\alpha}} \sum_{i,j=0}^{d-1} p_{i,j}|\epsilon_i^\alpha \rangle \langle \epsilon_j^\alpha| \vec{e}_\lambda(b^\dagger_\alpha + b_\alpha) \notag \\
	&+ \frac{i \alpha q}{\varepsilon_0} \sqrt{\frac{\omega_\alpha}{2}} \sum_{i,j=0}^{d-1} r_{i,j}|\epsilon_i^\alpha \rangle \langle \epsilon_j^\alpha| \vec{e}_\lambda(b^\dagger_\alpha - b_\alpha) \label{afQRM}.
\end{align}
Under the finite-level truncation, the gauge invariance is generally broken \cite{stokes2019gauge}. Setting $n=2$ leads to the QRM in the arbitrary-gauge formalism 
\begin{align}
	H^{\alpha}_{2} &= P_2^\alpha H P_2^\alpha \notag \\
	&= \frac{\Omega}{2}\sigma_z +  \omega_\alpha (b_\alpha^\dagger b_\alpha + \frac{1}{2}) + \Delta^\alpha \notag \\
	&- \frac{(1-\alpha)q}{n} \sqrt{\frac{1}{2\omega_\alpha}} \Omega r_{1,0}\sigma_y \vec{e}_\lambda(b^\dagger_\alpha + b_\alpha) \notag \\
	&+ \frac{i \alpha q}{\varepsilon_0} \sqrt{\frac{\omega_\alpha}{2}} \sigma_x r_{1,0} \vec{e}_\lambda(b^\dagger_\alpha - b_\alpha) \label{a2QRM}.
\end{align}
Here $\Omega=(\epsilon_1 - \varepsilon_0)$ is the transition frequency of the atom and $\Delta^\alpha = \frac{(\epsilon_1 + \varepsilon_0)}{2} + \frac{\alpha^2 q^2}{2v\varepsilon_0}(P_2^\alpha \vec{r} P_2^\alpha)^2$ is the energy shift. In the Coulomb gauge with $\alpha=0$, Eq.\eqref{a2QRM} restores the widely used standard QRM in most previous studies. Be aware that the derivation in Eq.\eqref{a2QRM} has assumed the symmetric potential about the origin, i.e. $\langle \epsilon_i^\alpha | \vec{r} | \epsilon_i^\alpha \rangle = 0$, and the resulted Hamiltonian \eqref{a2QRM} is simplified. The result in Eq.\eqref{a2QRM} is consistent with the one obtained in \cite{stokes2019gauge}. Furthermore, the QRM can be extended to any finite level-truncation conventionally in this framework.

\section{Numerical Results}\label{sec3}
We first consider the ground-state energy $E_g$, which is one of the most significant properties of the QRM. When the truncated dimension $d$, approaches a sufficiently large value close to infinity, the $d$-level QRM, $H_d^\alpha$, approximates the full Hamiltonian $H_\infty^\alpha$. In this case, the accurate ground-state energy, denoted as $E_{g,\infty}$, can be calculated using the expression $E_{g,\infty} = \langle\psi_\infty^{\alpha} | H_\infty^\alpha | \psi_\infty^{\alpha} \rangle$, where $| \psi_\infty^{\alpha} \rangle$ represents the precise ground state of the full Hamiltonian. On the other hand, for the finite-level QRM, the ground-state energy is calculated by $E_{g,d} = \langle \psi_d^{\alpha} | H_\infty^\alpha | \psi_d^{\alpha} \rangle$. Here $| \psi_d^{\alpha} \rangle$ corresponds to the ground state of $H_d^\alpha$. By comparing the predictions for $E_g$, we can determine the optimal gauge, in which the predictions of the finite-level QRM most closely match those of the full Hamiltonian.

Next, we examine two different potentials, namely the square-well potential \eqref{square_well_potential} and the harmonic oscillator potential \eqref{oscillator_potential}. The square-well potential is defined as follows:
\begin{align}
	V(x) =
	\begin{cases}
		0, & -1/2<x<1/2, \\
		\infty, & \text{elsewhere},
	\end{cases}\label{square_well_potential}
\end{align}
On the other hand, the harmonic oscillator potential is given by
\begin{align}
	V(x) = \frac{1}{2}x^2. \label{oscillator_potential}
\end{align}
Through the following calculation, the atomic frequency $\Omega$ of the two lowest levels in respective potentials is set as $\Omega=1$.

\begin{figure}[htp]
	\hspace*{-10mm}
	\includegraphics[scale=0.4]{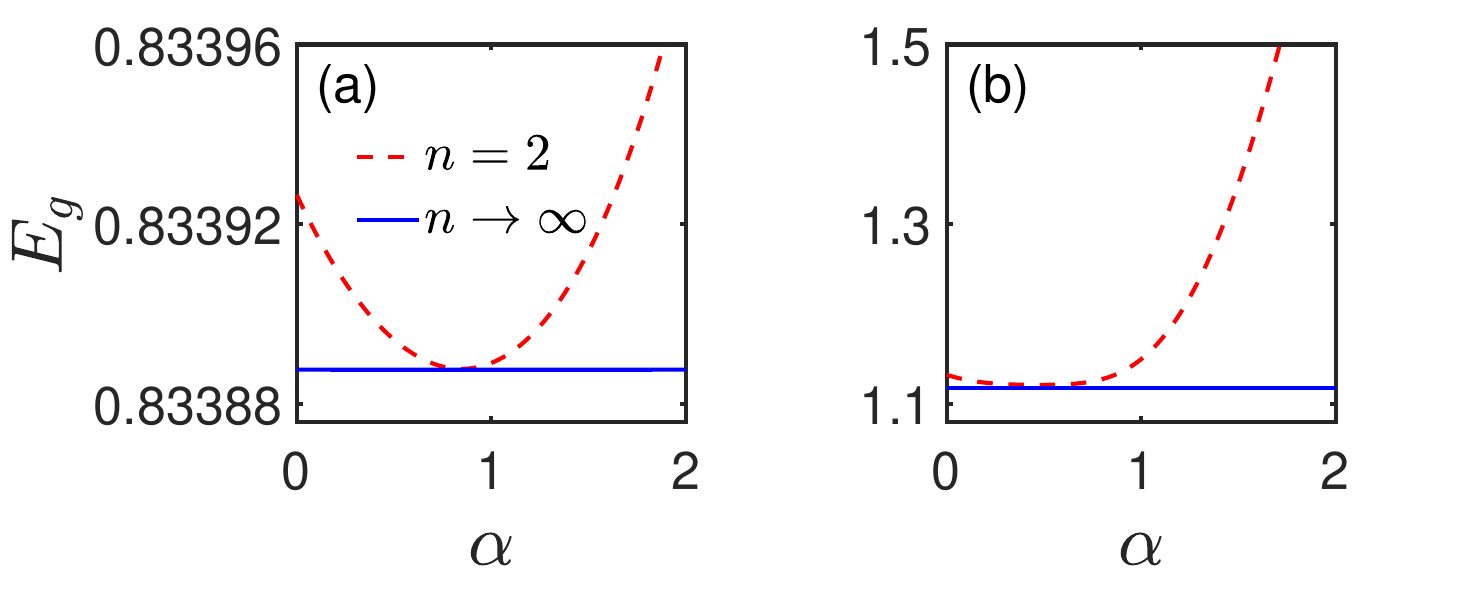}
	\caption{Determining the optimal gauge by finding the gauge that minimizes the difference between ground state energies of the two-level model and the full Hamiltonian with $\omega=1$. (a) In square-well potential, the optimal gauge is determined as $\alpha_o=0.847$. (b) In oscillator potential, the optimal gauge is determined as $\alpha_o=0.469$.}
	\label{fig1}
\end{figure}

To exhibit the gauge-dependent behavior of the ground state energy, we present a comparison of numerical in Fig.\ref{fig1}. The optimal gauge, denoted as $\alpha_o$, is identified as the gauge where the difference between the result of two-level model and the accurate ones is minimized. In Fig.\ref{fig1}(a), the comparison is conducted in the square-well potential. Notably, the optimal gauge $\alpha_o$ is found to be a value close to the dipole gauge ($\alpha=1$), specifically $\alpha_o=0.847$. We will later show that the optimal gauge generally does not takes the value $\alpha=1$, except for certain limiting cases. In Fig.\ref{fig1}(b), we observe that the optimal gauge deviates significantly from the dipole gauge for the oscillator potential, which is determined to be $\alpha_o=0.469$.

\begin{figure}[htp]
	\hspace*{-15mm}
	\includegraphics[scale=0.45]{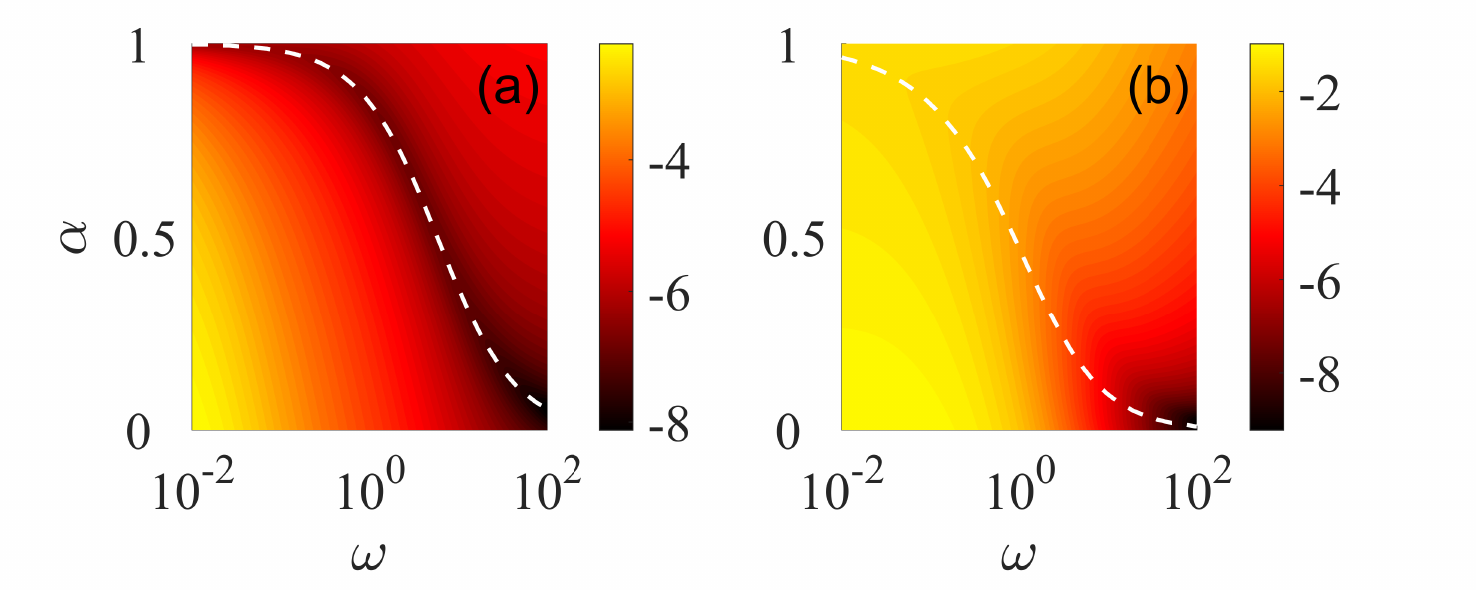}
	\caption{The optimal gauge is determined by finding the minimal value of $\log{\Delta E_g}$ under different $\omega$. The minimal values are marked by white dashed curves in (a) square-well potential and (b) oscillator potential. Asymptotic behaviours of the optimal gauge in terms of $\omega$ for both potentials can be extracted. In the limit of $\omega \to 0$, the optimal gauge takes $\alpha_o \to 1$, namely the dipole gauge. In contract, when $\omega \to \infty$, the optimal gauge takes $\alpha_o \to 0$, namely the Coulomb gauge.}
	\label{fig2}
\end{figure}

To further investigate the properties of the ground state energy, we analyze the dependence on both the gauge and the cavity frequency. For convenience, we define the difference between the two-level result and the accurate result as $\Delta E_g = E_{g,2} - E_{g,\infty}$. The optimal gauge is determined by minimizing $\Delta (E_g)$. In Fig.\ref{fig2}, we present a contour map of $\log{\Delta (E_g)}$. It is evident that for both potentials, the optimal gauge generally depends on the cavity frequency. In the limit of $\omega \to 0$, the optimal gauge takes the dipole gauge, while in the limit of $\omega \to \infty$, the optimal gauge restores the Coulomb gauge. A semi-classical picture may help us understand this result. In the limit of $\omega \to \infty$, the electromagnetic field undergoes rapid variations, preventing the atom's dipolar field from effectively responding to it. Consequently, the dipolar moment term in Eq.\eqref{full_H_a}, represented as $\alpha q\vec{r}$, becomes negligible. Conversely, when the frequency $\omega \to 0$, the photon's momentum diminishes, and the dipolar moment becomes significant. In this case, it is appropriate to choose the optimal gauge as $\alpha_o=1$.

\begin{figure}[htp]
	\hspace*{-5mm}
	\includegraphics[scale=0.4]{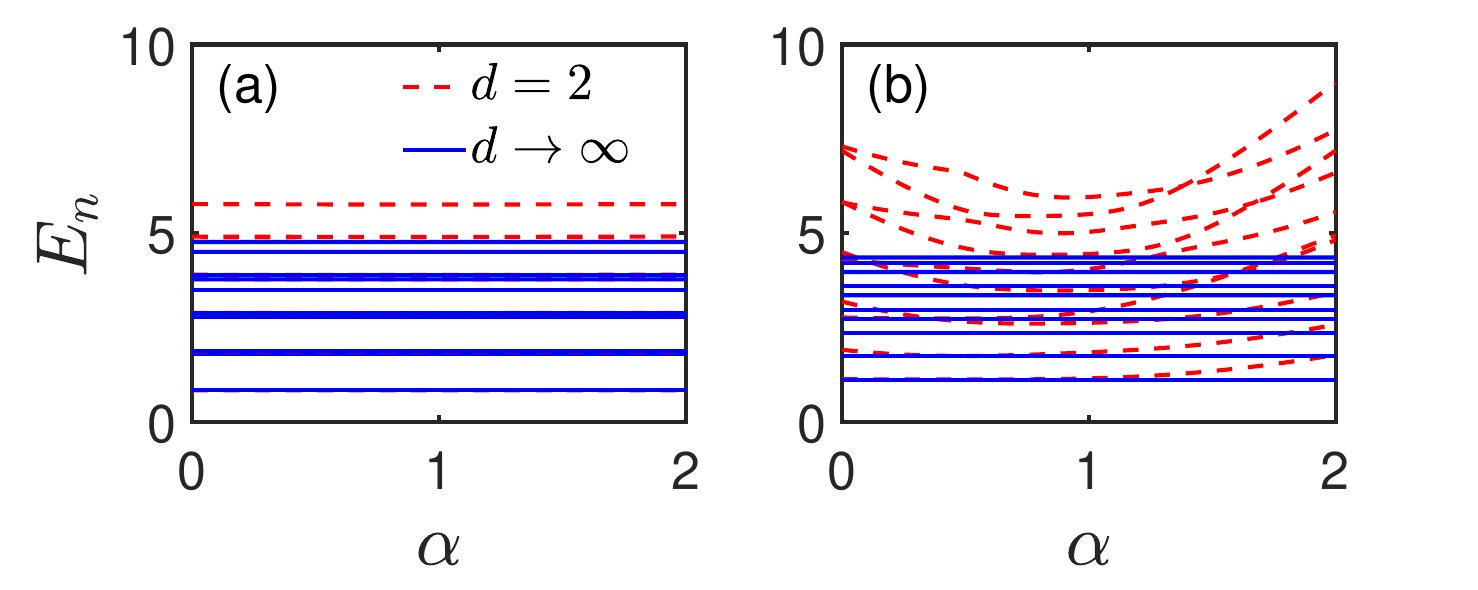}
	\caption{Behaviors of the energy spectrum of the first ten levels with $\omega=1$. (a) In square-well potential, the optimal gauges for each level are actually slightly different and close to $\alpha=1$. (b) In oscillator potential, the optimal gauges for each level are apparently different, and the degeneracy at certain gauges for high levels are observed.}
	\label{fig3}
\end{figure}

\begin{figure}[htp]
    \vspace*{-15mm}
	\hspace*{-5mm}
	\includegraphics[scale=0.43]{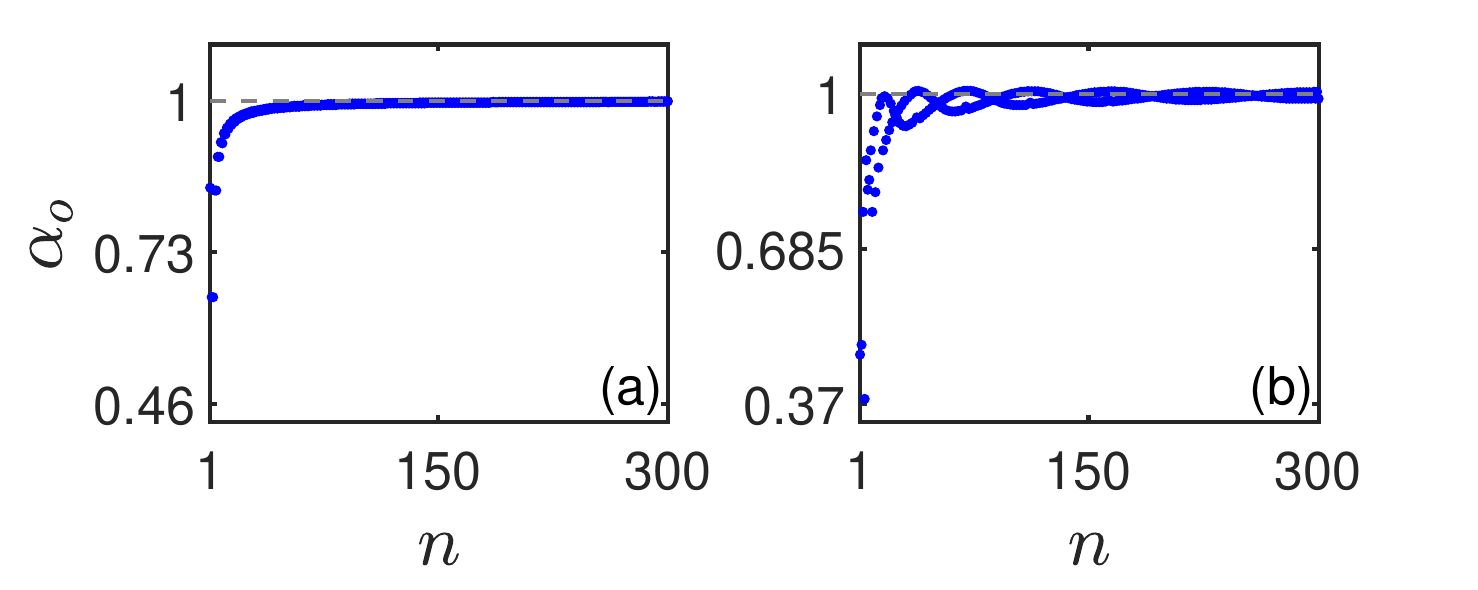}
	\caption{Behaviors of the optimal gauge in terms of system level $n$ with $\omega=1$ are discerned. When $n \to \infty$, the optimal gauge, though being different for lower levels, tends to $\alpha=1$. In addition, the behaviors of optimal gauges exhibit two distinct trajectories, which relate to the truncation dimension of atomic Hamiltonian. (a) In square-well potential, the optimal gauge choices tends to $\alpha=1$ in a monotonic way. (b) In oscillator potential, the optimal gauge choice converges in a  fluctuating manner.}
	\label{fig4}
\end{figure}

Another important quantity to consider is the energy spectrum of the system. In Fig.\ref{fig3}, we display the energy spectrum, indicated by $E_n$, of the first ten levels. However, the optimal gauge for each level is different as shown in Fig.\ref{fig4}(a), where $n$ denotes the $n$th $E_n$ level. For the higher levels, the optimal gauge asymptotically approaches the dipole gauge in an interesting way, which will be depicted later. In the case of the oscillator potential in Fig.\ref{fig3}(b), the phenomenon where no constant optimal gauge exists for the entire spectrum becomes more apparent. The behavior of the optimal gauges with respect to the energy levels is similar with the one for the square well potential, as shown in Fig.\ref{fig4}(b). For the higher levels, values of the optimal gauge converges to $\alpha=1$. A semi-classical picture may also shed light on understanding the asymptotic behaviours. When considering a highly excited state, the energy gap, or equivalently, the momentum of the fermion excitation, is significantly larger compared to the photon momentum. Thus the dipole is significant and $\alpha=1$ should be taken as the optimal gauge.

\begin{figure}[htp]
    \vspace*{-5mm}
	\hspace*{-15mm}
	\includegraphics[scale=0.43]{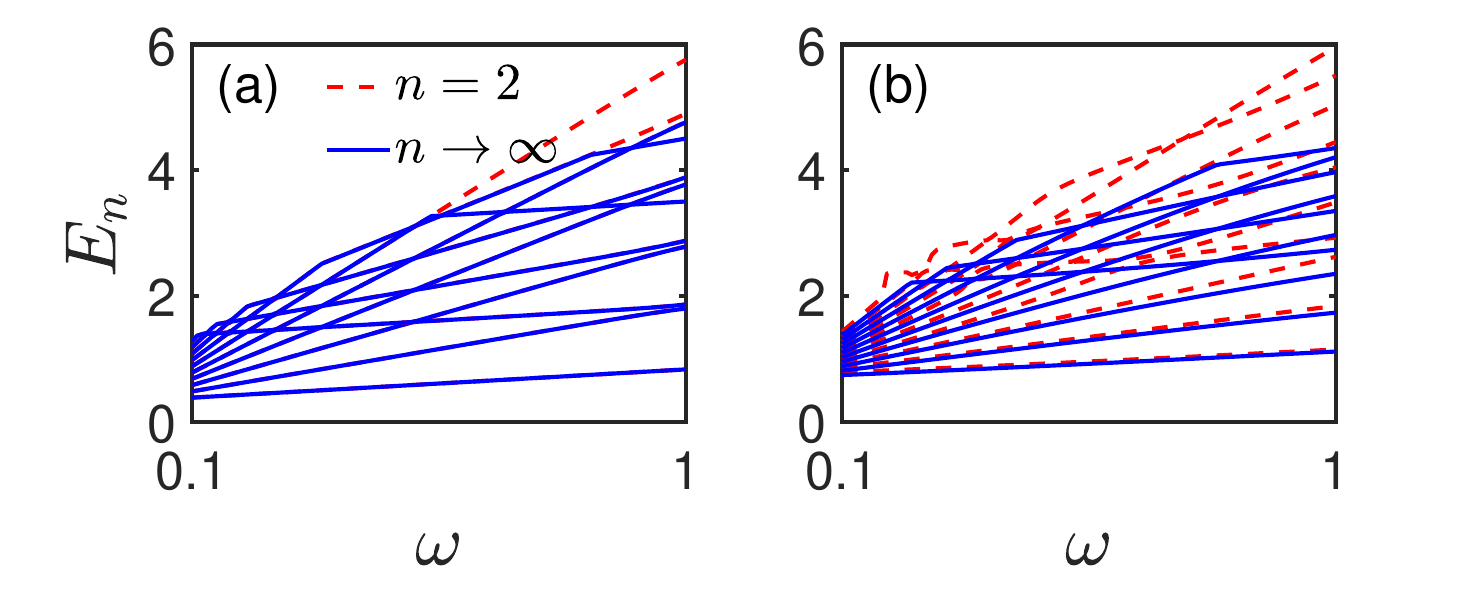}
	\caption{The dependence of $E_n$ on $\omega$ under $\alpha=1$. (a) The square well potential. (b) The oscillator potential.}
	\label{fig5}
\end{figure}

We also investigate the dependence of $E_n$ on $\omega$. In Fig.\ref{fig5}, we take $\alpha=1$ for both the square well potential (a) and the oscillator potential (b). When $\omega$ is small, the two-level results are close to the accurate ones, consistent with the conclusion drawn in Fig.\ref{fig2}, which indicates that when $\omega$ is small the optimal gauge always takes the value $\alpha=1$. When $\omega$ is large, the situation becomes complicated. As depicted in Fig.\ref{fig2}, the optimal gauge gradually approaches to $\alpha=0$. However, according to Fig.\ref{fig4}, for higher level the optimal gauge approaches $\alpha=1$. There seems to be a contrast between these two findings. Nonetheless, it is essential to note that for high-level spectra, as illustrated in Fig.\ref{fig3}, the two-level model fails to reproduce accurate results in any gauges. The optimal gauge is solely the choice that minimizes errors.

To investigate the dynamical behavior of the QRM, we calculate the out-time-order correlator (OTOC) in the equilibrium ground state. The OTOC is a crucial quantity for describing quantum chaos, many-body localization, and detecting quantum phase transitions \cite{sun2020out,rozenbaum2017lyapunov,shen2017out,swingle2018unscrambling}. The OTOC is defined as follows:
\begin{equation}
	F_{\text{eq}}(t) = \langle \psi_d^\alpha | W^\dagger(t) V^\dagger(0) W(t) V(0) | \psi_d^\alpha \rangle,
\end{equation}
where we choose $W = V = b_\alpha^\dagger b_\alpha$, the photon number operator, and $W(t) = e^{iH_\infty^\alpha t}W(0)e^{-iH_\infty^\alpha t}$.

\begin{figure}[htp]
        \vspace*{0mm}
	\hspace*{-5mm}\includegraphics[scale=0.43]{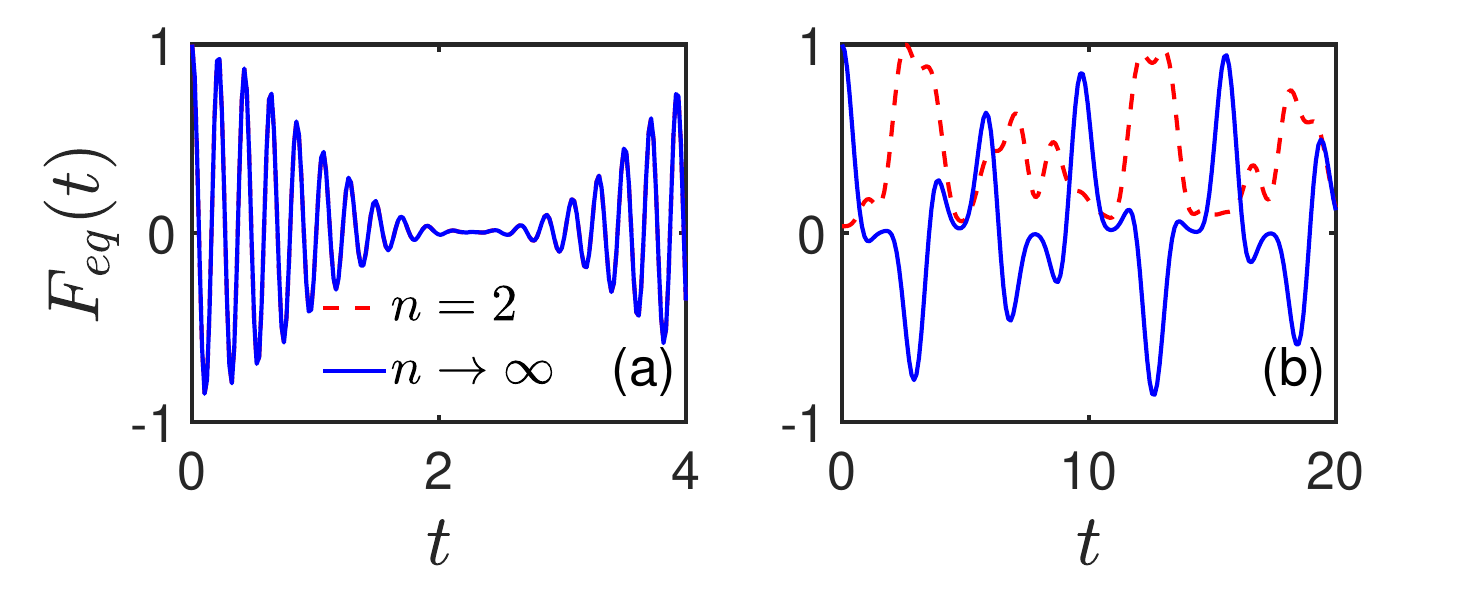}
	\caption{The time evolution of the OTOC in optimal static gauges of two potentials, namely (a) the square-well potential with $\alpha=0.987$ and $\omega=1$ and (b) the oscillator potential with $\alpha=0.469$ and $\omega=1$.}
	\label{fig6}
\end{figure}

\begin{figure}[htp]
    \vspace*{0mm}
	\hspace*{-5mm}
	\includegraphics[scale=0.43]{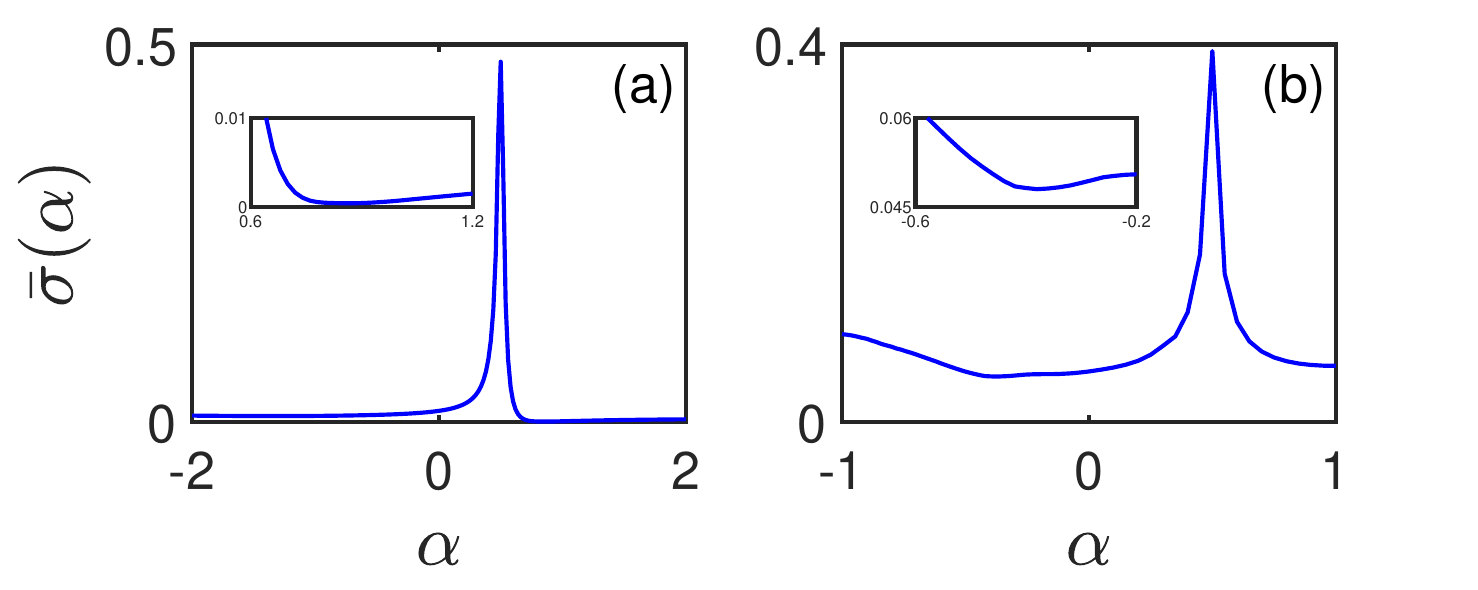}
	\caption{Determine the dynamical gauge by finding the minimal value of the mean error of the two-level OTOC compared to the accurate one. (a) In the square-well potential with $\omega=1$, the optimal dynamical gauge is $\bar{\alpha}_o=0.862$. (b) In oscillator potential with $\omega=1$, the optimal dynamical gauge is $\bar{\alpha}_o=-0.374$.}
	\label{fig7}
\end{figure}

\begin{figure}[htp]
    \vspace*{0mm}
	\hspace*{-15mm}
	\includegraphics[scale=0.43]{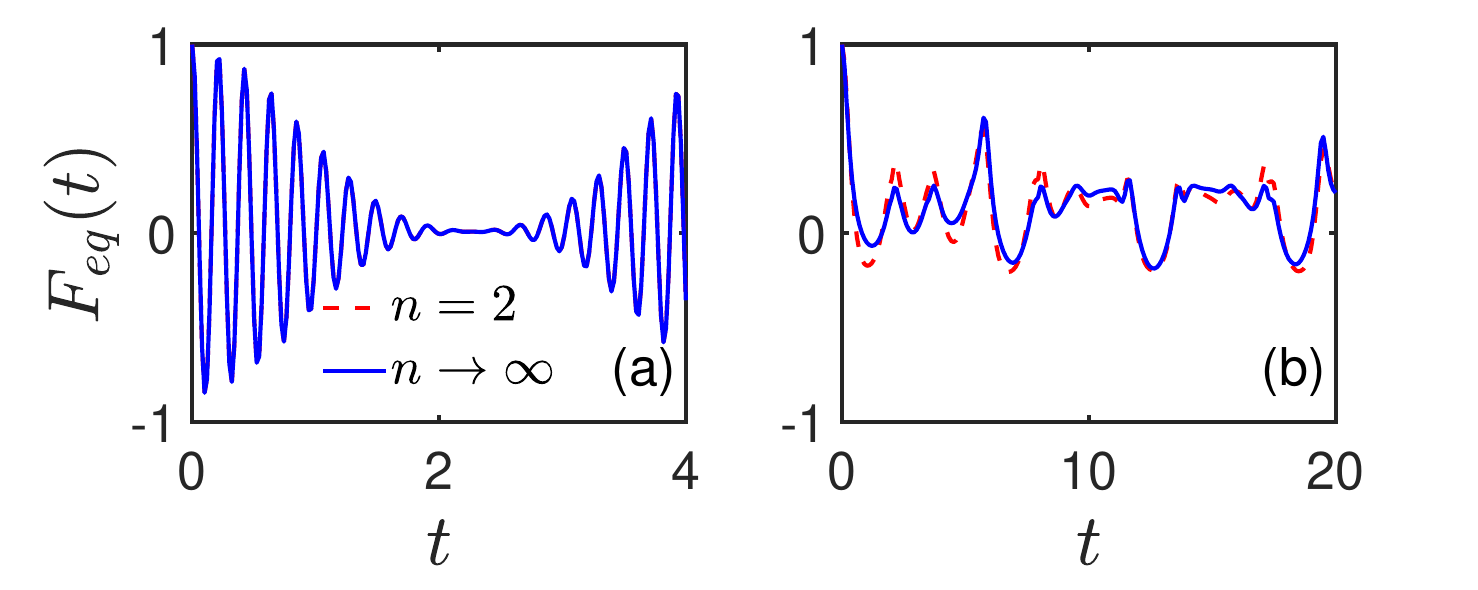}
	\caption{The time evolution of the OTOC in optimal dynamical gauges of two potentials, namely (a) the square-well potential with $\bar{\alpha}_o=0.862$ and $\omega=1$ and (b) the oscillator potential with $\bar{\alpha}_o=-0.374$ and $\omega=1$.}
	\label{fig8}
\end{figure}

\begin{figure}[htp]
    \vspace*{0mm}
	\hspace*{-15mm}
	\includegraphics[scale=0.43]{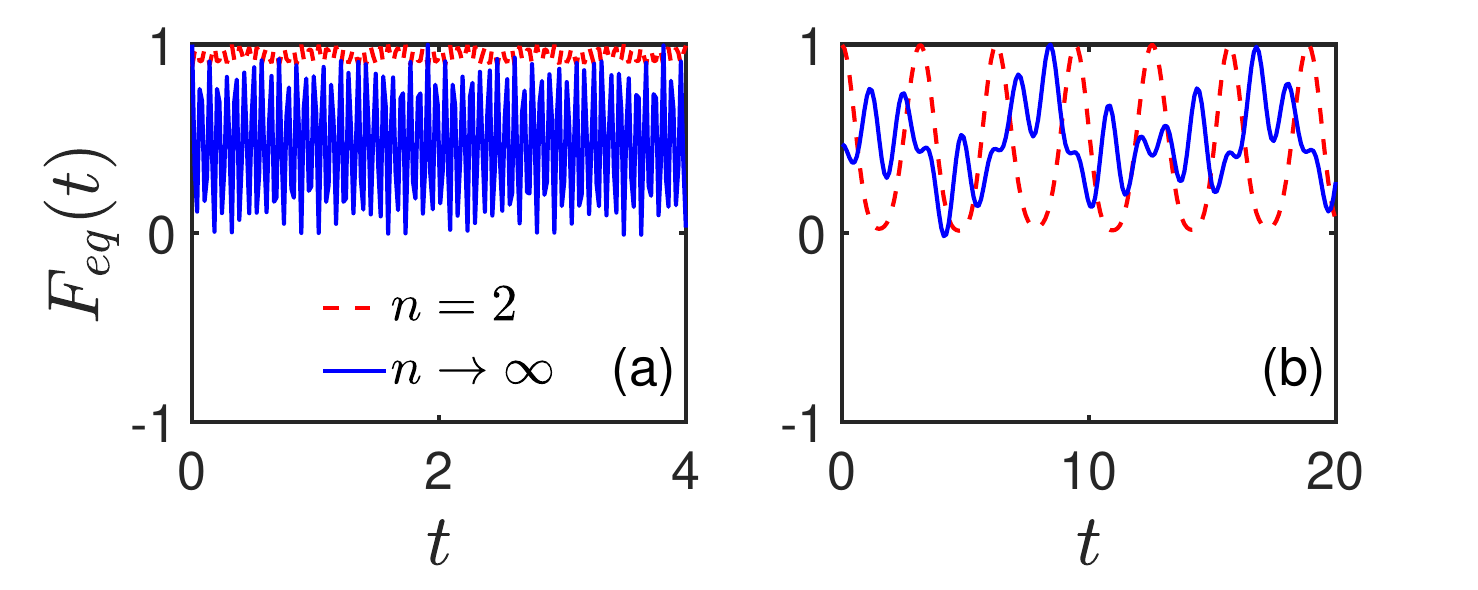}
	\caption{The time evolution of the OTOC in optimal dynamical gauges of two potentials, namely (a) the square-well potential with $\alpha=0.5$ and $\omega=1$ and (b) the oscillator potential with $\alpha=0.5$ and $\omega=1$.}
	\label{fig9}
\end{figure}

Figure \ref{fig6} presents the time evolution of the normalized OTOC in different potentials. In Fig.\ref{fig6}(a), we observe consistency between the two-level results and the accurate ones with the optimal static gauge $\alpha_o=0.987$ in the square well potential. However, we will later show that this gauge choice may not be the optimal one for OTOC. In Fig.\ref{fig6}(b), in the optimal static gauge of the oscillator potential, there is an apparent discrepancy between the two-level result and the accurate one, thereby suggesting an optimal dynamical gauge, denoted by $\bar{\alpha}_o$, is necessary. To determine $\bar{\alpha}_o$ for the OTOC, we define the mean error of the two-level OTOC compared to the accurate one as
\begin{equation}
    \bar{\sigma} = \frac{1}{T} \int_0^T |F_{\text{eq},2}(t)-F_{\text{eq},\infty}(t)| dt, \label{mean_error_of_OTOC}
\end{equation}
where $F_{\text{eq},2}$ denotes the two-level OTOC and $F_{\text{eq},\infty}$ denotes the accurate one. In Fig.\ref{fig7}, the dependence of $\bar{\sigma}$ on $\alpha$ is shown. We determine the optimal dynamical gauge of the OTOC in the square well potential as $\bar{\alpha}_o=0.862$ and in the oscillator potential as $\bar{\alpha}_o=-0.374$.
In addition, in the optimal dynamical gauge, the time evolution of OTOC is shown in Fig.\ref{fig8}, from which we observe that the discrepancy between the two-level OTOC and the accurate one is smaller than in the optimal static gauge, particularly for the oscillator potential.
Interestingly, we can find that, for both potentials, selecting $\alpha=0.5$ as the gauge choice for OTOC yields poorest results. To show the behavior of OTOC in this gauge, we present the time evolution in Fig.\ref{fig9}. In Fig.\ref{fig9}(a), which corresponds to the square well potential, the OTOC of the two-level model demonstrates slight fluctuations below $F_{eq}=1$. This suggests that the OTOC of the photon is weakly affected by the atomic transition, indicating a relatively effective weak coupling between the atom and the cavity mode. In Fig.\ref{fig9}(b), representing the oscillator potential, the OTOC of the two-level model exhibits simple harmonic behavior but with poor accuracy compared to the full-model counterparts.

\section{Conclusion\label{sec4}}
In the work, we investigate both the optimal static and dynamical gauge choice in the quantum Rabi model (QRM) in different potentials. For the optimal static gauge, we calculate and compare the two-level ground-state energy and low-energy spectra with the accurate ones. The optimal static gauge can be determined by finding the gauge choice where two-level results are most close to the accurate ones. We also reveal that the dipole gauge and the Coulomb gauge becomes the optimal gauges respectively in the low-frequency and high-frequency limit. In addition, the dipole gauge is also the optimal choice for higher levels in both the square well potential and the oscillator potential. We calculate the time evolution of the out-time-order correlator (OTOC) in the equilibrium ground state for both the two-level and the full Hamiltonian. The optimal dynamical gauge can be determined by finding the gauge choice where the minimal mean error between the two-level results and the accurate ones. The optimal dynamical gauges for both potentials differ from their static counterparts, emphasizing the necessity of introducing optimal dynamical gauges.

Establishing an analytical relationship between the optimal gauge and other influencing factors is crucial for developing a gauge-independent theory in a truncated Hilbert space, which is one of the most important task in the field of the light-matter interaction. Utilizing the gauge choice, in other words, the additional symmetries, contributes to construct accurate approximate theories for practical calculations. Semi-classical pictures in this work may shed light on the further analytical work. The framework employed in our work allows for straightforward extensions to any level truncation. It will be instrumental in future systematic and analytical studies addressing the gauge issue. The optimal gauge issue is not limited to the QRM but exists in any system, where redundant degrees of freedom are introduced to construct an effective theory. We look forward to further research that combines the idea of the optimal gauge with other more realistic systems.

\appendix

\section{Identification of the longitudinal component in the inverse space \label{appendixA}}
The freedom to choose $\vec{A}_\parallel$ in a QED theory is called gauge freedom. In the Lagrangian formalism as we will show, this gauge freedom is embodied in a total derivative term containing all the $\vec{A}_\parallel$ terms. The Lagrangian in Eq.\eqref{L_minimal_coupling} does not fix a gauge, i.e., $\vec{A}_\parallel$ is still arbitrary. To identify the $\vec{A}_\parallel$ terms, we first transfer to inverse space via a Fourier transform:
\begin{align}
	L &= \frac{1}{2}n \dot{\vec{r}}^2 - V_{ext}(\vec{r}) + \int_{half} d^3 k \{ \varepsilon_0 [|\dot{\vec{\tilde{A}}} + i\vec{k}\tilde{\phi}|^2 \notag \\  
	&-|\vec{k}\times \vec{\tilde{A}}|^2] + [\vec{\tilde{J}}^* \cdot \vec{\tilde{A}} + \vec{\tilde{J}}\cdot \vec{\tilde{A}}^* - \tilde{\rho}^* \tilde{\phi} - \tilde{\rho}\tilde{\phi}^*] \}. \label{L_in_inverse_space}
\end{align}
Here $\vec{J}(\vec{x})=q\dot{\vec{r}}\delta(\vec{x}-\vec{r})$ and $\phi(\vec{x})=q\delta(\vec{x}-\vec{r})$ are the current density and charge density, respectively, while $\vec{\tilde{J}}(\vec{k})$ and $\tilde{\phi}(\vec{k})$ are the corresponding Fourier transforms. We use the Parseval-Plancherel identity for vector fields $\vec{f}(\vec{x})$ and $\vec{g}(\vec{x})$:
\begin{align}
	\int d^3x [\vec{f}^*(\vec{x}) \cdot \vec{g}(\vec{x})] = \int d^3k [\vec{\tilde{f}}^*(\vec{k}) \cdot \vec{\tilde{g}}^*(\vec{k})]
\end{align}
The subscript $half$ in Eq.\eqref{L_in_inverse_space} denotes integration over half-space where $\vec{k}_{\lambda} > 0$. This is possible because the physical fields $\vec{E}(\vec{x}),\vec{B}(\vec{x})$ are real. For real $\vec{E}(\vec{x})$, $\vec{E}^*(\vec{k}) = \vec{E}(-\vec{k})$ and:
\begin{align}
	\int d^3k \vec{E}^*(\vec{k}) \cdot \vec{E}(\vec{k}) = 2\int_{half} d^3k \vec{E}^*(\vec{k}) \cdot \vec{E}(\vec{k})
\end{align}
Observing the divergence identity of $\vec{E}(\vec{x})$ in inverse space: 
\begin{align}
	\tilde{\phi} = \frac{1}{k^2}\left[ ik\dot{\vec{\tilde{A}}}_\parallel+\frac{\tilde{\rho}}{\varepsilon_0} \right] \label{E_divergence_in_inverse_space}
\end{align}
We see that both the $\phi$ and $\vec{A}$ terms contain $\vec{A}_{\parallel}$. Substituting Eq. \eqref{E_divergence_in_inverse_space} into Eq. \eqref{L_in_inverse_space} and regrouping all $\vec{A}_\parallel$ terms gives:
\begin{align}
	L &= \frac{1}{2} n\dot{\vec{r}}^{2} + \int_{half} d^3k ( -\frac{\tilde{\rho}^* \tilde{\rho}}{\varepsilon_0 k^2} ) \notag \\  
	&+ \int_{half} d^3k \{ \tilde{\vec{J}}_\perp \cdot \tilde{\vec{A}}_\perp^* + \tilde{\vec{J}}_\perp^* \cdot \tilde{\vec{A}}_\perp + \varepsilon_0 [\dot{\tilde{\vec{A}}}^*_\perp \cdot \dot{\tilde{\vec{A}}}_\perp -  k^2 \tilde{\vec{A}}^*_\perp \cdot \tilde{\vec{A}}_\perp] \} \notag \\
	&+ \int_{half} d^3k (- \frac{i\tilde{\rho}^*\dot{\tilde{A}}_\parallel}{k} + \frac{i\tilde{\rho}\dot{\tilde{A}}^*_\parallel}{k}+\tilde{\vec{J}}_\parallel^* \cdot \tilde{\vec{A}}_\parallel+\tilde{\vec{J}}_\parallel\cdot \tilde{\vec{A}}_\parallel^*). \label{L_in_inverse_space_regroup}
\end{align}
The $\vec{A}_\parallel$ term in Eq. \eqref{L_in_inverse_space_regroup} is a total time derivative:
\begin{align}
	&\int_{half} d^3k (- \frac{i\tilde{\rho}^*\dot{\tilde{A}}_\parallel}{k} + \frac{i\tilde{\rho}\dot{\tilde{A}}^*_\parallel}{k}+\tilde{\vec{J}}_\parallel^* \cdot \tilde{\vec{A}}_\parallel+\tilde{\vec{J}}_\parallel\cdot \tilde{\vec{A}}_\parallel^*) \\  
	=& \int d^3 k [\frac{i}{k} \frac{d}{dt}(\tilde{\rho} \tilde{A}_\parallel^* - \tilde{\rho}^*\tilde{A}_\parallel)]. \label{total_derivative}
\end{align}
We used charge conservation in inverse space: $\dot{\tilde{\rho}} = -i\vec{k} \cdot \tilde{\vec{{J}}}_\parallel$. This shows $\vec{A}_\parallel$ is redundant as it is in a total time derivative term, which does not affect the equations of motion. In the Coulomb gauge, we set $\vec{A}_\parallel=0$, so the total derivative vanishes and we get Eq. \eqref{L_C}.

\bibliography{ref}
	
\end{document}